\documentclass[12pt,preprint]{aastex}

\newcommand{\chandra}{\mbox{\em Chandra\/}}
\newcommand{\asca}{{\it ASCA\/}}

\shorttitle{X-ray Source Catalog of the ECDFS}
\shortauthors{Virani et al.}

\begin{document}

\title{The Extended \chandra\ Deep Field-South Survey: X-ray Point-Source Catalog}

\author{Shanil N. Virani, Ezequiel Treister\altaffilmark{1,2}, C. Megan Urry\altaffilmark{1}, and Eric Gawiser\altaffilmark{1,3}}
\affil{Department of Astronomy, Yale University, P.O.~Box 208101,
    New Haven, CT 06520}

\altaffiltext{1}{Yale Center for Astronomy and Astrophysics, Yale University, P.O.~Box 208121, New Haven, CT 06520}
\altaffiltext{2}{Departamento de Astronomia, Universidad de Chile, Casilla 36-D, Santiago, Chile.}
\altaffiltext{3}{NSF Astronomy and Astrophysics Postdoctoral Fellow}
\email{svirani@astro.yale.edu}
\slugcomment{Submitted to AJ}

\begin{abstract}

The Extended \chandra\ Deep Field-South (ECDFS) survey
consists of 4 \chandra\ ACIS-I pointings and covers $\approx$ 1100 square 
arcminutes ($\approx$ 0.3 deg$^2$) centered on the original CDF-S field
to a depth
of approximately 228 ks. This is the largest \chandra\ survey ever conducted 
at such depth, and only one XMM-Newton survey reaches a lower flux limit in
the hard 2.0--8.0 keV band.
We detect 651 unique sources --- 587 using a conservative source detection 
threshold and 64 using a lower source detection threshold. These are presented 
as two separate catalogs. Of the 651 total sources, 561 are detected in
the full 0.5--8.0 keV band, 529 in the soft 0.5--2.0 keV band, and 335
in the hard 2.0--8.0 keV band. 
For point sources near the
aim point, the limiting fluxes are approximately $1.7 \times
10^{-16}$ $\rm{erg~cm^{-2}~s^{-1}}$ and $3.9 \times
10^{-16}$ $\rm{erg~cm^{-2}~s^{-1}}$ in the 0.5--2.0 keV and
2.0--8.0 keV bands, respectively. Using simulations, we determine the catalog
completeness as a function of flux and assess uncertainties in the derived
fluxes due to incomplete spectral information. We present the differential 
and cumulative flux distributions, which are in good agreement
with the number counts from previous deep X-ray surveys and with the
predictions from an AGN population synthesis model that can
explain the X-ray background. In general, fainter sources
have harder X-ray spectra, consistent with the hypothesis that
these sources are mainly obscured AGN.
\end{abstract}


\section{Introduction}

Wide-area X-ray surveys have played a fundamental role in
understanding the nature of the sources that populate the
X-ray universe. Early surveys like the \textit{Einstein}
Medium Sensitivity Survey \citep{gioia90}, the ROSAT
International X-ray/Optical Survey \citep{ciliegi97}, and
the ASCA Large Sky Survey \citep{akiyama00} showed that the
vast majority of the bright X-ray sources are active galactic
nuclei (AGNs). More specifically, shallow wide-area surveys
in the soft (0.5--2.0 keV) X-ray band yield mostly unobscured, 
broad-line AGNs, which are
characterized by a soft X-ray spectrum with a photon index
of $\Gamma \simeq$ 1.9 \citep{nandra94}. In contrast, deep X-ray 
surveys --- particularly surveys that make use of the unprecedented, 
sub-arsecond spatial resolution of the \chandra\ X-ray Observatory ---
find AGN with harder X-ray spectra ($\Gamma \sim$ 1.4) at fainter fluxes,
more like the hard spectrum of the X-ray background.

Deep \chandra\ surveys have thus opened a new vista on resolving 
the X-ray background and identifying the role and evolution of 
accretion power in all galaxies. The cosmic X-ray
background is now almost completely resolved ($\sim$
70--90\%) into discrete sources in the deep, pencil beam
surveys like the \chandra\ Deep Fields (CDF-N, \citealp{brandt01}; 
CDF-S, \citealp{giacconi02}). To understand the composition
of the sources that make up the X-ray background,
population synthesis models have been constructed
\citep{madau94,comastri95,gilli99,gilli01,treister05} which typically require
approximately 3 times as many obscured AGN as traditional Type 1 (unobscured)
AGN.


While the deep fields provide the deepest view of the X-ray
universe and have generated plentiful AGN samples at lower luminosities, 
the small area covered by pencil-beam surveys means luminous sources are
poorly sampled. In an attempt to determine the
luminosity function of X-ray emitting AGN up to $z \sim 5$,
as well as to leverage existing deep multiwavelength data in the extended
$30' \times 30'$ field centered on the CDF-S, the region
surrounding the CDF-S was recently observed by $Chandra$. Covering
$\approx$ 1100 square arcminutes ($\approx$ 0.3 deg$^2$), the Extended
\chandra\ Deep Field-South (ECDFS) survey is the
largest \chandra\ survey field at this depth ($\approx$ 230 ks), and is the 
second deepest and widest survey ever conducted in the X-rays (the XMM-Newton
survey of the Lockman Hole  is deeper in the
hard band and has $\sim$ 30\% more area; \citealt{hasinger04}).

In this paper, we present the X-ray catalog for the ECDFS and the
number counts in two energy bands. In subsequent papers,
we will present the optical and near-IR properties of these X-ray sources,
including first results from our deep optical spectroscopy campaign 
obtained as part of the one-square-degree MUltiwavelength Survey by 
Yale/Chile (MUSYC) \citep{gawiser05}.

In Section 2, we describe our data reduction procedure. 
In Section 3, we describe the point source detection and astrometry.
The X-ray source catalog and basic survey results are presented in 
Section 4 and the conclusions are given in Section 5. The average Galactic
column density along this line of sight for the four pointings
is $9.0 \times
10^{19}$~cm$^{-2}$ \citep{stark92}. $H_0=70$~km~s$^{-1}$
Mpc$^{-1}$, $\Omega_{\rm M}=0.3$, and $\Omega_{\Lambda}=0.7$
are adopted throughout this paper which is consistent with
the cosmological parameters reported by
\citet{spergel03}.
All coordinates throughout this paper are J2000.

\section{Observations and Data Reduction}

\subsection{Instrumentation and Diary of Observations}

All nine observations of the ECDFS survey field were
conducted with the Advanced CCD Imaging Spectrometer (ACIS)
on-board the \chandra\ X-ray Observatory\footnote{For
additional information on the ACIS and \chandra\, see the
\chandra\ Proposers' Guide at
http://cxc.harvard.edu/proposer/POG/html.} as part of the approved
guest observer program in Cycle 5 (proposal number 05900218 -- PI Niel 
Brandt; \citealt{lehmer05}). 
ACIS consists of 10 CCDs, distributed in a 2$\times$2 array (ACIS-I) and a
1$\times$6 array (ACIS-S). All 4 of the ACIS-I CCDs are
front-illuminated (FI) CCDs; 2 of the 6 ACIS-S CCDs are
back-illuminated CCDs (S1 and S3). Of these 10 CCDs, at most
6 can be operated at any one
time. Table~\ref{tbl-diary} presents a journal of the
\chandra\ observations of the ECDFS. All nine observations
were conducted in VERY FAINT mode (See the \chandra\
Proposers' Guide, pg. 95) so that the pixel values of the
$5\times5$ event island are telemetered rather than just the
$3\times3$ event island as in FAINT mode. This telemetry
format offers the advantage of further reducing the
instrument background after ground processing (see Section
2.2). Observation ids (ObsIds) 5019--5022 and 6164
also had the ACIS-S2 CCD powered on (see
Table~\ref{tbl-diary}). However, due to the large off-axis
angle of the S2 CCD during these observations, it has a much
broader point spread function (PSF) and hence lower sensitivity 
so we exclude data from this CCD and focus only on data collected
from the ACIS-I CCDs. The on-axis CCD for each
ACIS-I observation is I3 and the ACIS-I field of view is
$16\farcm 9\times 16\farcm 9$.

\subsection{Data Reduction}

All data were re-processed using the latest version of the
\chandra\ Interactive Analysis of Observations\footnote{See
http://cxc.harvard.edu/ciao/.} (CIAO; Version 3.2.1;
released 10 February 2005) software as well as version 3.0.0
of the calibration database (CALDB; released 12 December
2004). We chose to re-reduce all nine datasets rather than
simply use the standard data processing (SDP) level 2 event
files because we wanted the datasets to be reduced in a
consistent manner, and more importantly, we wanted to take
advantage of better \chandra\ X-ray Center (CXC) algorithms to 
reduce the ACIS
particle background by using the information
located in the outer 16 pixels of the $5\times5$ event
island\footnote{See ``Reducing ACIS Quiescent Background
Using Very Faint Mode'',
http://cxc.harvard.edu/cal/Acis/Cal\_prods/vfbkgrnd/index.html.},
as well as executing a new script that is much more
efficient at identifying ACIS ``hot pixels'' and cosmic ray
afterglow events\footnote{See
http://cxc.harvard.edu/ciao/threads/acishotpixels/.}. This new
hot pixel and cosmic ray afterglow
tool is now implemented in the standard data
processing pipeline at the CXC but was not applied in the SDP pipeline
of the present observations.

The following procedure was used to arrive at a new level
2 event file. Before applying the new CIAO tool to
identify ACIS hot pixels and cosmic ray afterglow
events, the pixels identified in the CXC-provided level 1
event file as being due to a cosmic ray afterglow event were
reset. An afterglow is the residual charge from the
interaction of a cosmic ray in a front-side illuminated CCD
frame. Some of the excess charge is captured by charge traps
created by the radiation damage suffered early in the
mission (see \citealp{townsely00} and
references therein) and released over the next few to a few dozen
frames. If these afterglow events are not
removed from the data, they can result in the spurious
detection of faint sources. To better account for such
events, a new, more precise method for identifying afterglow
events was developed by the CXC and has now been introduced
into the standard data processing pipeline. This CIAO tool,
``acis\_run\_hotpix,'' was then run on the reset level 1
event file to identify and flag hot pixels and afterglow
events in all 9 ACIS observations of the ECDFS. The last
step in producing the new level 2 event file was to run the
CIAO tool ``acis\_process\_events.'' In addition to applying
the newest gain map supplied in the latest release of the
CALDB, this tool also applies the pixel randomization and
the ACIS charge transfer inefficiency (CTI) correction. The
latter corrects the data for radiation damage sustained by
the CCDs early in the mission. All of these corrections are part of
the standard data processing and are on by default in
acis\_process\_events. The time-dependent gain correction is
also applied to the event list to correct the
pulse-invariant (PI) energy channel for the secular drift in
the average pulse-height amplitude (PHA) values. This drift is 
caused primarily by
gradual degradation of the CTI of the ACIS CCDs (e.g.,
\citealp{Schwartz04}). Finally, the observation-specific
bad pixel map created by ``acis\_run\_hotpix'' was supplied,
and the option to clean the ACIS particle background by
making use of the additional pixels telemetered in VERY
FAINT mode was turned on.

Once a new level 2 event file was produced, we applied the
standard grade filtering to each observation, choosing only
event grades 0, 2, 3, 4, and 6 (the standard \asca\ grade
set), and the standard Good Time Intervals supplied by
the SDP pipeline.  We also restrict the energy range to
0.5--8.0 keV, as the background rises steeply below and
above those limits\footnote{See
http://cxc.harvard.edu/contrib/maxim/stowed/.}.  Lastly, we
examined the background light curves for all 9 observations
as the ACIS background is known to vary significantly. For
example, \citet{plucinsky00} found that the
front-illuminated CCDs can show typical increases of 1 - 5
cts s$^{-1}$ above the quiescent level, while the
back-illuminated CCDs can show large excursions --- as high
as 100 cts s$^{-1}$ above the quiescent level --- during
background flares. The durations of these intervals of enhanced
background are highly variable, ranging from 500 s to
10$^4$ s.  The cause of these background flares is currently
not known (see \citealp{grant02}); however, they may be caused by
low-energy protons ($<$100 keV; e.g., \citealp{plucinsky00,struder01}). 
The time periods corresponding to these background flares are
generally excised from the data before proceeding with
further analysis although not always (see
\citealp{brandt01,nandra05}). In fact, \citet{dwkim04a}
find that the source detection probability depends strongly
on the background rate.  To examine our observations for
such periods, we used the \rm{CIAO} script
\rm{ANALYSE\_LTCRV.SL} which identifies periods where the
background is $\pm$3$\sigma$ above the mean. All 9
observations were filtered according to this
prescription (see Table~\ref{tbl-diary} for a comparison of
raw exposure time vs. filtered exposure time), resulting in only 
$\sim$40.6 ks ($\sim$4\%)
being lost due to background flares (954.2 ks vs. 913.6
ks). Of this, 20.3 ks were excluded from the end of ObsID 5017
due to a flare in which the count rate increased by a
factor of $\sim$2. Table~\ref{table-exp} lists the net
exposure time for each of the 4 pointings used to image the
ECDFS region. The net exposure time for each of the 4
pointings varies from a low of 205 ks to a high of 239 ks,
with the mean net exposure time for the entire survey field
of 228 ks. These extra steps in processing help remove spurious sources 
and result in fewer catalog sources than if the standard processing or 
pipeline products were used.

\section{Data Analysis}

In this paper, we report on the sources detected in three
standard X-ray bands (see Table~\ref{table-defn}): 0.5--8.0 keV 
(full band), 0.5--2.0 keV (soft band), and 2.0--8.0 keV (hard band). 
The raw ACIS resolution is 0.492 arcsec pixel$^{-1}$, however, source
detection and flux determinations were performed on the
block 4 images, \textit{i.e.,} 1.964 arcsec pixel$^{-1}$, as the
source detect tool and exposure map generation 
require significant computer resources for full size images; for greater
accuracy, source positions were determined from the block 1 images.

\subsection{Image and Exposure Map Creation}

\begin{figure}
\epsscale{1.0}
\plotone{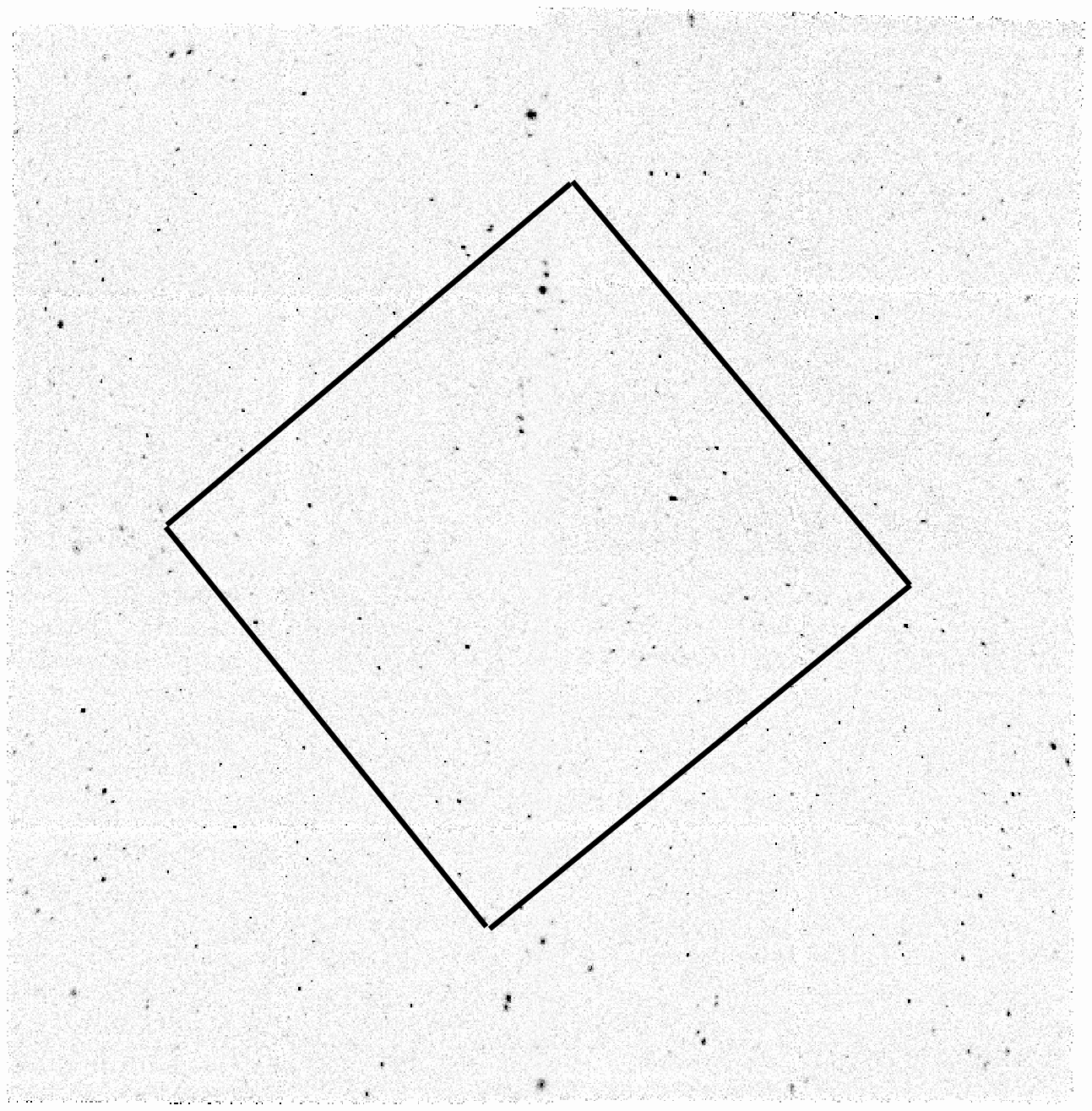}
\caption{Exposure-corrected full band (0.5--8.0 keV) image of the 
ECDFS. This image has been binned by a factor of four in both 
RA and Dec, and has been made using the standard \asca\ grade set. 
The black square superimposed on the raw image is the 
approximate footprint (most of the exposure lies within this region)
of the CDF-S proper \citep{giacconi02}.\label{full}}
\end{figure}

Observations at each of the four pointings were combined via
the CIAO script ``merge\_all.'' This script was executed using
CIAO version 2.3 because of a known bug in the ``asphist''
tool under CIAO version 3.2.1; this bug results in incorrect
exposure maps for the merged image\footnote{See the usage
warning at
http://cxc.harvard.edu/ciao/threads/merge\_all/.}. At each
pointing, the observation with the longest integration time
was used for co-ordinate registration. For example, when
merging ObsIds 5015 and 5016, the merged event list was
registered to ObsId 5015 as it has approximately twice the
integration time as 5016. Table~\ref{table-exp} lists the
ObsIds for each pointing, as well as the raw and the net integration time.
For each pointing, we constructed images in the three standard
bands: 0.5--8.0 keV (full band), 0.5--2.0 keV (soft
band), and 2.0--8.0 keV (hard band); see
Table~\ref{table-defn}. The full band exposure-corrected
image for the entire survey field\footnote{Raw
and smoothed \asca\ -grade images for all three
standard bands (See Table~\ref{table-defn}) are available
from http://www.astro.yale.edu/svirani/ecdfs/.}
is presented in Figure~\ref{full}.

We constructed exposure maps in these three energy bands for
each pointing and for the entire survey field\footnote{Exposure maps for 
all three standard bands (see Table~\ref{table-defn}) are available
from http://www.astro.yale.edu/svirani/ecdfs/.}. 
These exposure maps were created in the standard way and are
normalized to the effective exposure of a source location at
the aim point. The procedure used to create these exposure
maps accounts for the effects of vignetting, gaps between
the CCDs, bad column filtering, and bad pixel
filtering. However, it should be noted that charge
blooms caused by cosmic rays can reduce the detector
efficiency by as much as few percent\footnote{See
http://cxc.harvard.edu/ciao/caveats/acis\_caveats\_050620.html.}. 
There is currently no
way to account for such charge cascades; however, when a tool
becomes available, we will correct for this effect as
necessary and make the new exposure maps publicly available at the
World Wide site listed in Footnote 7. 
The exposure maps
were binned by 4 so that they were congruent to the final
reduced images. A photon index of $\Gamma$~=~1.4,
the slope of the X-ray background in the 0.5--8.0 keV band
(e.g. \citealp{marshall80,gendreau95,kushino02}) was used in
creating these exposure maps. 

\begin{figure}
\epsscale{1.0}
\plotone{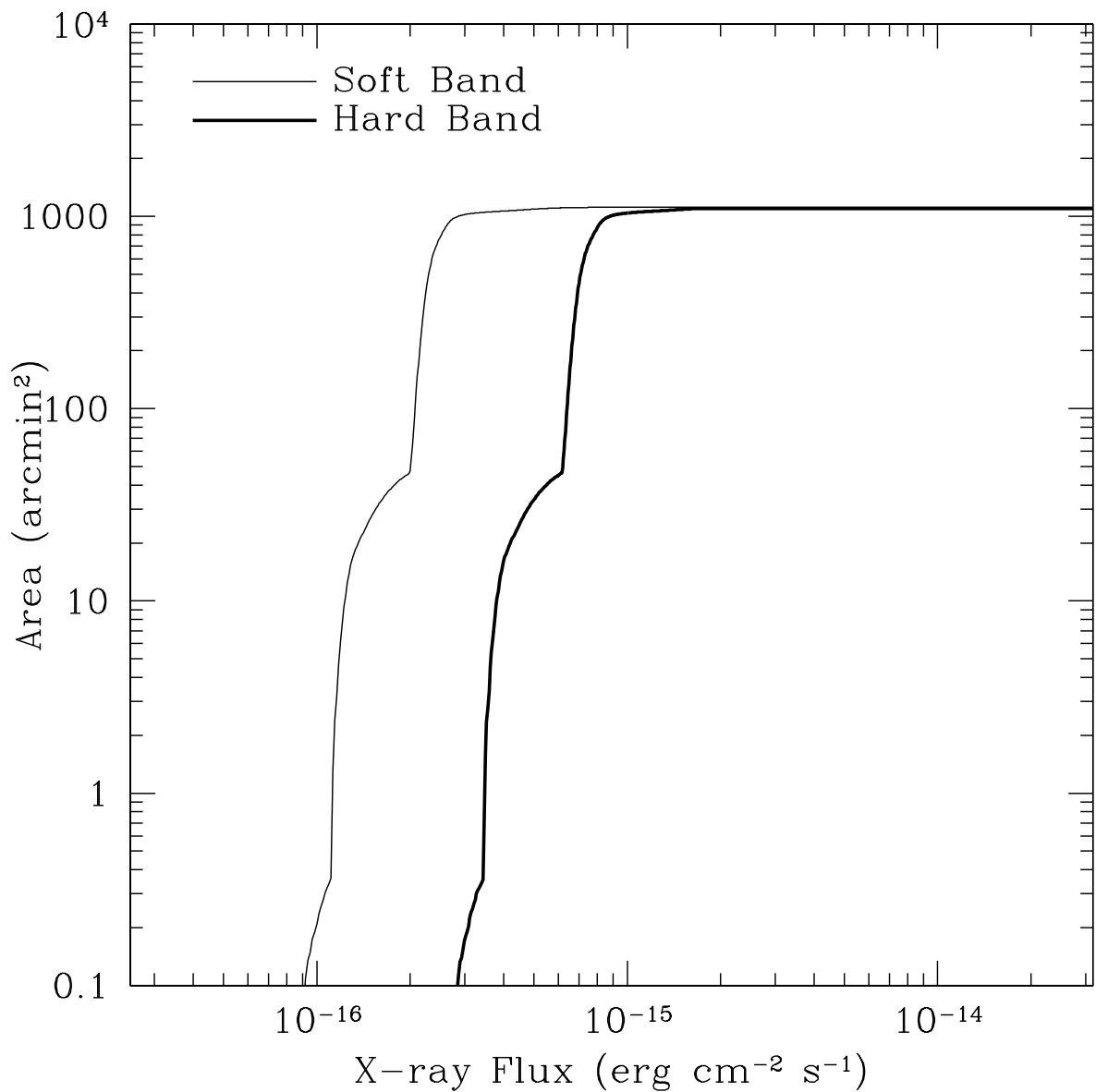}
\caption{The survey area vs. limiting flux for the two bands for which we have
calculated the log $N$--log $S$ function: soft band
(\textit{thin line}) and the hard band (\textit{thick line}). The total area of
the survey is $\approx$ 1100 arcmin$^2$
($\sim$0.3~deg$^2$).\label{area}}
\end{figure}

In order to calculate the survey area as a function of the X-ray flux 
in the soft and hard bands, we used the exposure maps generated for each band
and assumed a fixed detection threshold of 5 counts in the soft band and 2.5 in
the hard band ($\sim$2$\sigma$). Dividing these counts by the exposure map,
we obtain the flux limit at each pixel for each band. The pixel area is then
converted into a solid angle and the cumulative histogram of the flux limit
is constructed (Figure~\ref{area}). The total survey area is 
$\approx$ 1100 arcmin$^2$ ($\approx$ 0.3 deg$^2$). A more precise method of
determining the survey area as a function of the X-ray flux is described
by \cite{kenter03}; however, this would affect only the faint tail of the 
sample and would not significantly alter the present results. Therefore,
a more sophisticated treatment is deferred to a later paper.

\subsection{Point Source Detection}

To perform X-ray source detection, we applied the CIAO
wavelet detection algorithm \textit{wavdetect}
\citep{freeman02}. Although several other methods have
been used in other survey fields to find sources in
\chandra\ observations (e.g.,
\citealp{giacconi02,nandra05}), we chose \textit{wavdetect} to
allow a straightforward comparison between sources found in
our catalog with those found in the CDF-S
\citep{giacconi02,alexander03}. Moreover, \textit{wavdetect}
is more robust in detecting individual sources in crowded
fields and in identifying extended sources than the other
CIAO detection algorithm, \textit{celldetect}. Point-source
detection was performed in each standard band (see
Table~\ref{table-defn}) using a ``$\sqrt{2}$~sequence'' of
wavelet scales; scales of 1, $\sqrt{2}$, 2, $2\sqrt{2}$, 4,
$4\sqrt{2}$, and 8 pixels were used. \citet{brandt01}, for
example, showed that using larger scales can detect a few
additional sources at large off-axis angles but found that
this ``$\sqrt{2}$~sequence'' gave the best overall
performance across the CDF-N field. Moreover, as
\citet{alexander03} point out, sources
found with larger scales tend to have source properties and
positions too poorly defined to give useful results.

Our criterion for source detection is that a source must be
found with a false-positive probability threshold
($p_{thresh}$) of $1\times 10^{-7}$ in at least one of the
three standard bands. This false-positive probability threshold
is typical for point-source catalogs
(e.g., \citealp{alexander03,wang04}), although
\citet{dwkim04a} found that a significance threshold
parameter of $1\times 10^{-6}$ gave the most efficient
results in the \chandra\ Multiwavelength Project (ChaMP) survey. 
We ran \textit{wavdetect} using both probability thresholds and found that
using the lower significance threshold (\textit{i.e.}, $1\times 10^{-6}$)
results in only an additional 64 unique sources. Visual inspection of each of
these sources suggest they are bona fide X-ray sources. However, because these 
are sources found with the lower significance threshold, we present them
in a separate table (the secondary catalog; Table~\ref{cat2}). 
The primary catalog (Table~\ref{cat}) is a compilation of 587 unique
sources found using the higher significance threshold in at least one of the 
three energy bands.
For the remaining source detection
parameters, we used the
default values specified in CIAO which included requiring
that a minimum of 10\% of the on-axis exposure was needed in
a pixel before proceeding to analyze it. We also
applied the exposure maps generated for each pointing 
(see Section 3.1) to mitigate finding
spurious sources which are most often located at the edge of
the field of view.

The number of spurious sources per pointing is approximately
$p_{thresh} \times N_{pix}$, where $N_{pix}$ is the total
number of pixels in the image, according to the
\textit{wavdetect} documentation. Since there are
approximately 2 $\times$ 10$^6$ pixels in each image for
each pointing, we expect $\sim$0.2 spurious sources per
pointing per band for a probability threshold of $1\times 10^{-7}$. 
Therefore, treating the 12 images
searched as independent, we expect $\sim$ 2-3 false sources
in our primary catalog (Table~\ref{cat}) for the case of a uniform
background. Of course the background is neither perfectly
uniform nor static as the level decreases in the gaps
between the CCDs and increases slightly near bright point
sources.  As mentioned by \citet{brandt01} and
\citet{alexander03}, one might expect the number of false
sources to be increased by a factor of $\sim$ 2--3 due to
the large variation in effective exposure time across the
field and the increase in background near bright sources due
to the point-spread function (PSF) wings. But our false-source estimate is
likely to be conservative by a similar factor
since \textit{wavdetect} suppresses
fluctuations on scales smaller than the PSF. That is, a
single pixel is unlikely to be considered a source detection
cell --- particularly at large off-axis angles \citep{alexander03}.

The source lists generated by the procedure above for each
of the standard bands in each of the pointings of the
ECDFS were merged to create the point-source catalogs
presented in Tables~\ref{cat} and~\ref{cat2}. The source positions listed in
each catalog are the full band \textit{wavdetect}-determined
positions except when the source was detected only in the
soft or hard bands. 
To identify the same source in the
different energy bands, a matching radius of 2.\arcsec5 or twice
the PSF size of each detect cell, whichever was the largest,
was used. For comparison, \citet{alexander03} and \citet{nandra05} used a
matching radius of 
2.\arcsec5 for sources within 6\arcmin ~of the aimpoint, and 4.\arcsec0 for
sources with larger off-axis angles. 
With our method, 9 and 3 soft- and hard-band 
sources, respectively, have more than 1 counterpart, so we took the 
closest one.
Note that both Tables ~\ref{cat} and ~\ref{cat2} excludes sources found
by \textit{wavdetect} in which one or both of the axes of
the ``source ellipse'' collapsed to zero. Over the survey field, 70 such 
sources are found; in general, these are unusual sources and although the 
formal probability of being spurious is low, there may be problems with these
detections. \citet{horn01}
found that using the \textit{wavdetect}-determined counts for
such objects as we do results in a gross underestimate of the number
of counts even though the source was detected with a
probability threshold of $1\times 10^{-7}$.
 Since these sources would appear in catalogs that do circular
aperture photometry instead, we present this list in a separate catalog 
(Table~\ref{cat3}) for completeness. 

Below we define the columns in Tables~\ref{cat} and~\ref{cat2}, our primary
and secondary source catalogs for the ECDFS survey.

\begin{itemize}

\item Column~1 gives the ID number of the source in our catalog. 

\item Column~2 indicates the International Astronomical Union approved names
for the sources in this catalog. All sources begin with the acronym ``CXOYECDF'' 
(for ``Yale E-CDF'')\footnote{Name registration submitted to http://cdsweb.u-strasbg.fr/viz-bin/DicForm.}.

\item Columns~3 and 4 give the right ascension and declination, respectively.
These are \textit{wavdetect}-determined positions for the unbinned images. If a source is detected in
multiple bands, then we quote the position determined in the full band; when a 
source is not detected in the full band, we quote the soft-band position or the
hard-band position.

\item Column~5 gives the PSF cell size, in units of arcseconds, as determined
by \textit{wavdetect}. The farther off-axis a source lies, the larger the 
PSF size.

\item Columns~6, 7, and 8 give the count rates (in units of cts~s$^{-1}$)
in the full band and the
corresponding upper and lower errors estimated  according to the
prescription of \citet{gehrels86}. If a source is undetected in this band, no
count rate is tabulated.

\item Columns~9, 10, and 11 give the count rates (in units of cts~s$^{-1}$)
in the soft band and the
corresponding upper and lower errors estimated  according to the
prescription of \citet{gehrels86}. If a source is undetected in this band, no
count rate is tabulated.

\item Columns~12, 13, and 14 give the count rates (in units of cts~s$^{-1}$)
in the hard band and the
corresponding upper and lower errors estimated  according to the
prescription of \citet{gehrels86}. If a source is undetected in this band, no
count rate is tabulated.

\item Column 15 lists the full band flux (in units of ~erg~cm$^{-2}$~s$^{-1}$) 
calculated using a photon slope of 
$\Gamma = $1.4 and corrected for Galactic absorption. If a source was undetected in the full band but was detected in the hard or soft band, the hard- or soft- band
flux (in that order of priority) was used to extrapolate to the full band assuming a photon slope of 1.4.

\item Column 16 lists the soft band flux (in units of ~erg~cm$^{-2}$~s$^{-1}$)
calculated using a photon slope of 
$\Gamma = $1.4 and corrected for Galactic absorption. If a source was undetected in the soft band but was detected in the full or hard band, the full- or hard- band
flux (in that order of priority) was used to extrapolate to the soft band 
assuming a photon slope of 1.4.

\item Column 17 lists the hard band flux (in units of ~erg~cm$^{-2}$~s$^{-1}$)
calculated using a photon slope of 
$\Gamma = $1.4 and corrected for Galactic absorption. If a source was undetected in the hard band but was detected in the full or soft band, the full- or soft- band
flux (in that order of priority) was used to extrapolate to the hard band 
assuming a photon slope of 1.4.

\item Column 18 provides individual notes for each source.
Examples include the
catalog ID (c\#) if detected in the CDF-S by \citet{alexander03}, or if the 
source was selected from a band other the full band ('h' or 's') or only 
detected in the full band ('f').

\end{itemize}

To determine source counts for each of our sources, we
extracted counts in the standard bands from each of the 
images using the geometry of the \textit{wavdetect}
source cell and the \textit{wavdetect}-determined source
position. For example, to determine the counts in the soft band, we used
the position and geometry determined by \textit{wavdetect} in the soft band
image to extract soft band counts. Some studies use circular aperature
photometry to extract sources counts. However, as both \citet{horn01} and 
\citeauthor{yang04} (\citeyear{yang04}; see their Figure 5) demonstrate, both techniques generally
return the same result.
Net count rates were then calculated using the
effective exposure (which includes vignetting) for each
pointing (exposure maps generated as described in Section 3.1). 
Errors were derived following \citet{gehrels86}, assuming an 84\% 
confidence level. Note that the exposure maps do account for the 
degradation of the soft
X-ray response of ACIS due to the build-up of a
contamination layer on the ACIS optical blocking filter
(\citealp{marshall04}; see Section 3.4). Therefore, the count rates
reported in Table~\ref{cat} are exposure- and
contamination-corrected. 

In Table~\ref{tbl-sourcesummary} we
summarize the source detections in the three standard bands,
and in Table~\ref{tbl-sourcesummary2} we summarize the
number of sources detected in one band but not in
another. To convert the count rates to flux, we determined
the conversion factor for each band assuming a photon
slope of $\Gamma$~=~1.4 and the mean Galactic $N_H$
absorption along the line-of-sight for each of the 4
pointings (N$_H$ = 9 $\times$ 10$^{19}$ cm$^{-2}$; \citealp{stark92}).

Our faintest soft-band sources have $\approx 4$ counts
(about one every 1.5 days), and our faintest hard-band
sources have $\approx 6$ counts; these sources are detected
near the aim point. The corresponding 0.5--2.0~keV and 2--8~keV 
flux limits, corrected for the Galactic column density, are 
$\approx 1.7\times
10^{-16}$~erg~cm$^{-2}$~s$^{-1}$ and $\approx 3.9\times
10^{-16}$~erg~cm$^{-2}$~s$^{-1}$, respectively. Of course,
these flux limits vary and generally increase across the field of view.

Undoubtedly, there are some sources in Table~\ref{cat} that are extended 
sources (\textit{i.e.}, resolved by \textit{Chandra}). \citet{giacconi02} find
18 extended sources in their 1 Ms catalog of the CDF-S 
out of 346 unique sources. The ECDFS survey has approximately 25\% the
integration time of the CDF-S but is approximately 3 times larger in area. 
Therefore, we expect roughly the same fraction of our sources reported in 
Table~\ref{cat} are likely to be extended. The identification, X-ray, and optical
properties of these sources will be presented in a later paper.

\subsection{Astrometry}

Given the superb \chandra\ spatial resolution, the on-axis
positional accuracy is often quoted as being accurate to
within 1\arcsec ~(e.g., \citealp{dwkim04a}); in fact, the 
overall 90\% uncertainty
circle of a \chandra\ X-ray absolute position has a radius of
0.6 arcsec, and the 99\% limit on positional accuracy is 0.8
arcsec\footnote{See http://cxc.harvard.edu/cal/ASPECT/celmon/.}. Nevertheless,
as the off-axis angle increases, the PSF broadens and
becomes circularly asymmetric (see \chandra\ Proposer's Guide; URL listed in 
Footnote 1). Therefore, source positions for faint sources at
large off-axis angles may not be accurate. In order to test
the astrometry of the \textit{wavdetect}-determined
positions, we have matched our full-band X-ray positions
provided in Table~\ref{cat} against deep $BVR$-band imaging
produced by the MUltiwavelength Survey by Yale/Chile
(MUSYC\footnote{For more information:
http://www.astro.yale.edu/musyc/.}; \citealp{gawiser05}). The
5$\sigma$ depth of the MUSYC optical imaging of this field
is 27.1 AB mag with approximately 0.\arcsec85 ~seeing. 
Correlating the X-ray positions reported in Tables~\ref{cat} and \ref{cat2}
with the optical positions found for sources in the ECDFS
field, we find that approximately 72\% of the sources reported in 
Table~\ref{cat} and 41\% of the sources reported in 
Table~\ref{cat2} have an optical counterpart within
1.\arcsec5 ~of the X-ray position. Furthermore, comparing the
X-ray positions with the optical positions for these matched
sources, we find a mean offset of -0.\arcsec08 ~in RA and +0.\arcsec28
~in Dec. (We do not correct the X-ray positions for
these offsets.) The optical properties of these X-ray
sources will be presented in a forthcoming paper (Virani et
al. 2005b, in prep.).

\subsection{Accuracy of Source Detections and Fluxes}

\begin{figure}
\plottwo{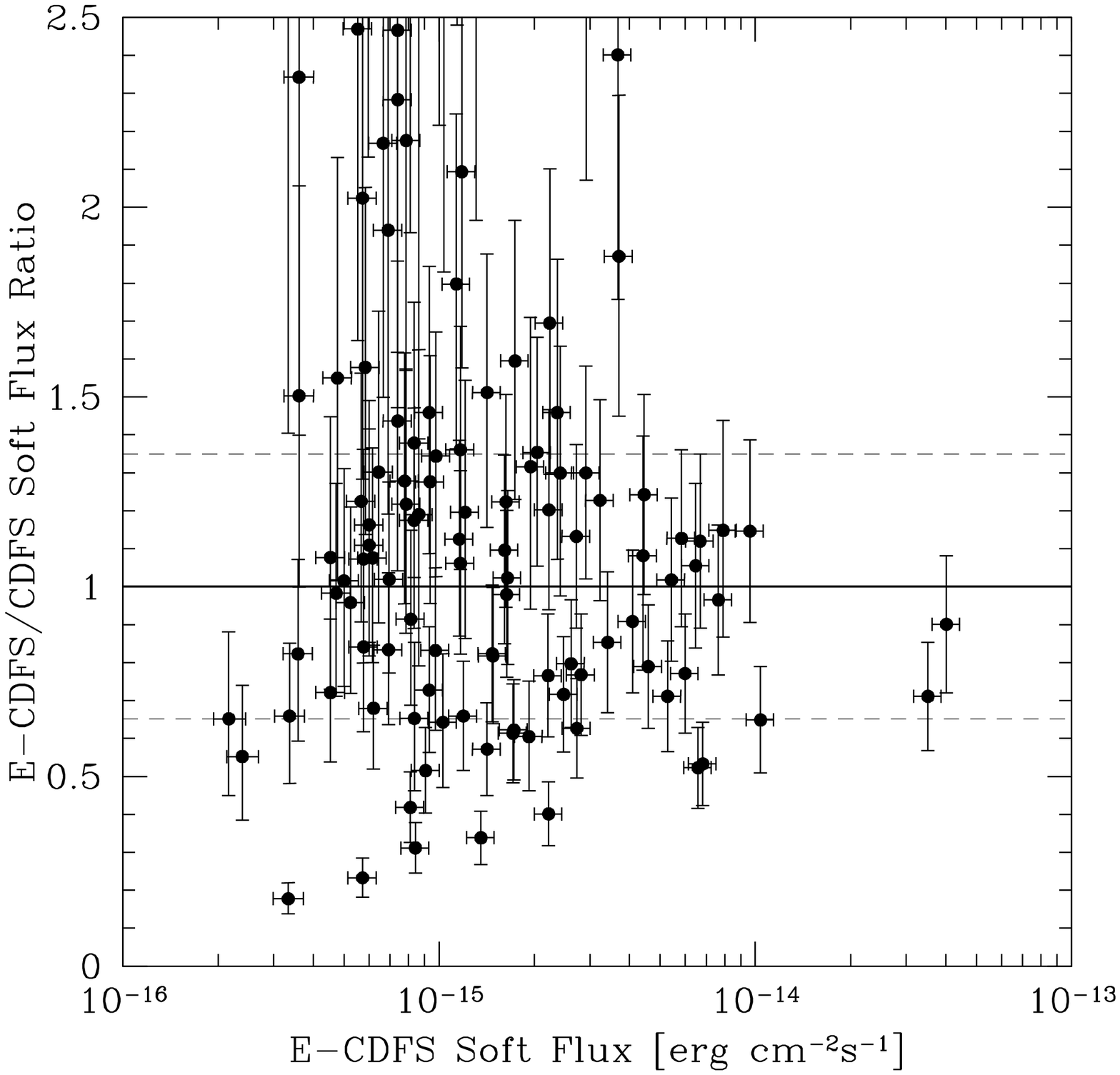}{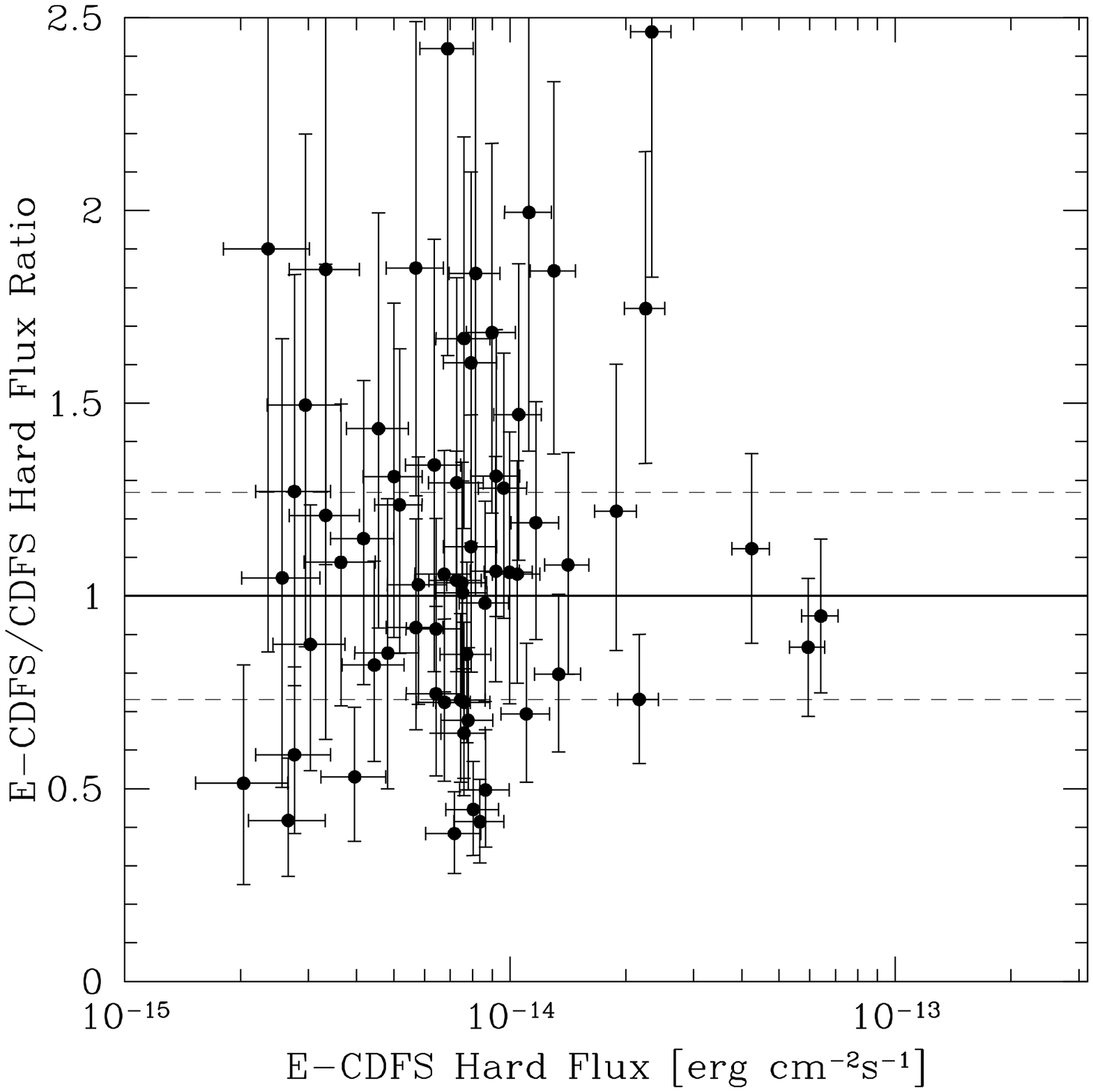}
\caption{Comparison of X-ray fluxes for the 115 sources (\textit{left panel}) and the
89 hard sources (\textit{right panel}) detected in the
CDF-S catalog of \citet{alexander03}. For this comparison, X-ray fluxes 
are not corrected for intrinsic Galactic absorption. In general, there is
very good agreement between the two independent data sets, with an
average flux ratio in the soft band of 1.14, with an RMS of 35\%,
shown by the dashed horizontal lines. In the hard band the
average flux ratio is 1.11 with an RMS of 27\%. These differences are 
explained by different treatments of the contamination layer and spectral 
slope, and suggest that
systematic uncertainties in the flux are $\approx$ 10-15\%. Note that some AGN 
may have actually varied between these two epochs. Error bars are calculated 
by adding in quadrature the statistical (Poisson) uncertainties in the counts 
plus a 10\% error arising from the likely range in spectral slopes (see 
Section 3.5). \label{scat}}
\end{figure}

Approximately one third of the ECDFS field overlaps with the 1 Ms
\chandra\ Deep Field South (see Figure~\ref{full} for the field layout). 
This is very useful as it allows us to
compare our results with the properties of the overlapping
sources already published. In particular, we used the
catalog of \citet{alexander03}, who re-analyzed the
original CDF-S data. In Figure~\ref{scat} we show the ratio of
our fluxes to those reported by \citet{alexander03} for the
overlapping sources. For this comparison, neither the CDF-S nor the 
ECDFS sources were corrected for intrinsic Galactic absorption.
(This correction
is $\simeq$4\% in the soft band and is negligible in the
hard band.) Error bars are calculated by adding in quadrature the statistical
(Poisson) uncertainties in the counts plus a 10\% error arising from the likely
range in spectral slopes (see Section 3.5).

Sources were matched using the closest CDF-S
counterpart to each ECDFS source, using a maximum search
radius of $\sim 2''$. To compare the fluxes of matched sources in the two
data sets, we excluded the most discrepant top and
bottom 15\% of the flux ratios, and found our fluxes are $\sim$14\% higher
in the soft band and $\sim$11\% higher in the hard band. In
the first case, the difference can be explained by the
different treatment of the contamination layer, which is
particularly important in the soft band. The
\citet{alexander03} catalog used ACISABS\footnote{Available
at
http://www.astro.psu.edu/users/chartas/xcontdir/xcont.html}
to correct their fluxes for the presence of a contamination
layer in the ACIS instrument. This tool assumes a spatially-uniform
contamination layer composed of hydrogen, carbon, nitrogen, and
oxygen. However, recent analysis of grating data
\citep{marshall04} shows that the amount of
contamination correction depends on the spatial position on the instrument,
and that the actual composition of the contamination is hydrogen, carbon,
oxygen, and fluorine (P. Plucinsky, priv. comm.). 
These two new discoveries may have caused
\citet{alexander03} to underestimate the contamination
correction, thus making their fluxes lower in the soft
band. In the hard band, the discrepancy can be explained by
our assumed value of $\Gamma =$ 1.4 for the spectral slope to calculate
fluxes, while \citet{alexander03} used individual spectral fits for
most of these overlapping sources. We conclude that the fluxes are broadly
consistent and that systematic uncertainties in their average values are
$\sim$ 15\%, although individual fluxes have larger uncertainties (and some
AGN may have actually varied).

\subsection{Simulations}

We performed extensive XSPEC and MARX simulations to investigate the 
statistical properties of the catalog, its completeness, and its flux limits.
First, in order to investigate the effect of a fixed photon
slope on the true flux of sources found in the ECDFS, we simulated 2000
sources with extreme photon spectral slopes, $\Gamma$=1 and $\Gamma$=2,
and with fluxes distributed smoothly from the minimum to the maximum in our
sample. We then computed their count rates in a typical ECDFS pointing 
($\sim$ 230 ks). Using a fixed photon slope of $\Gamma$=1.4 to compute fluxes 
then results in systematic
flux errors of $\sim$ 10\% in both the hard and soft bands. 

To investigate the completeness of our catalog, we used MARX to simulate X-ray 
images of sources with known properties, including the range of count rates
from just below our threshold to just above our highest count rate, and a
generous range of spectral slopes (1 $\leq \Gamma \leq$ 2) drawn from the observed
$\Gamma$ distribution observed in the 1 Ms CDFS survey \citep{alexander03}.
We positioned 1000 sources of known fluxes (consistent with an 
exposure time of $\sim$ 230 ks) randomly within the ECDFS survey field, so
the background and noise properties of the data are real. 
We then analyzed these simulated data with the same procedures used on the
real ECDFS data; that is, we performed source detection on the resulting event list 
via \textit{wavdetect}. This resulted in $\sim$90\% of the sources being 
recovered overall, with incompleteness becoming important below
$\sim$2$\times 10^{-16}$~erg~cm$^{-2}$~s$^{-1}$ 
and
$\sim$2$\times 10^{-15}$~erg~cm$^{-2}$~s$^{-1}$,
in the soft and hard bands, respectively.

\section{Results and Discussion}

We found 651 unique sources in the Extended \chandra\ Deep Field-South survey
field, which spans $\approx$ 0.3 deg$^2$ on the sky. Of these, 561 were 
detected in the 0.5--8.0 keV full band,
529 in the 0.5--2.0 keV soft band, and 335 in the 2.0--8.0 keV hard band. 
There are 9 hard-band sources that are not detected in either the soft or
full bands, 81 soft-band sources are not detected in either the hard or
full bands, and 56 full-band sources are not detected in either the 
soft or hard bands (see Table~\ref{tbl-sourcesummary2}). Of the 335 hard-band sources, 83 were not detected in the 
soft band ($\sim$20\%); these are candidates for highly absorbed sources. 
Of the 529 and 335 sources detected in the soft and hard bands, respectively, 
118 and 73 are detected in the CDF-S itself. Over this 0.11 deg$^2$ area, with
an exposure time of $\sim$ 1 Ms, \citet{giacconi02} found 346 unique sources, 
of which
307 were detected in the 0.5--2.0 keV band and 251 in the 2--10 keV band. 
In the CDF-N, with an area similar to the CDF-S but with twice the exposure, 
\citet{alexander03} found 503 X-ray sources in the 2 Ms exposure. The number
of sources found in the ECDFS is consistent with these two pencil
beam surveys, given an approximate slope of unity for the X-ray counts in this
flux range.

\begin{figure}
\plottwo{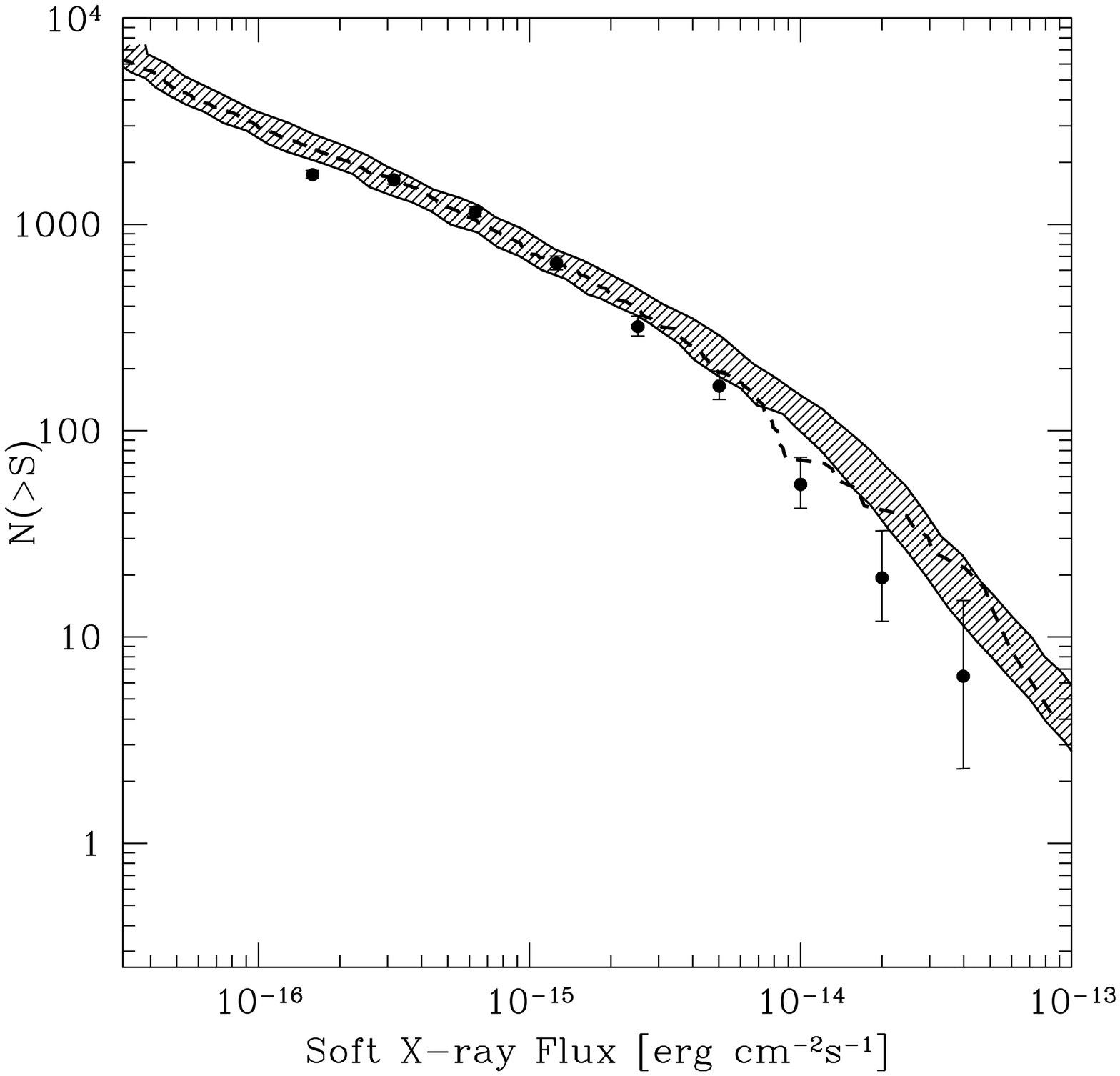}{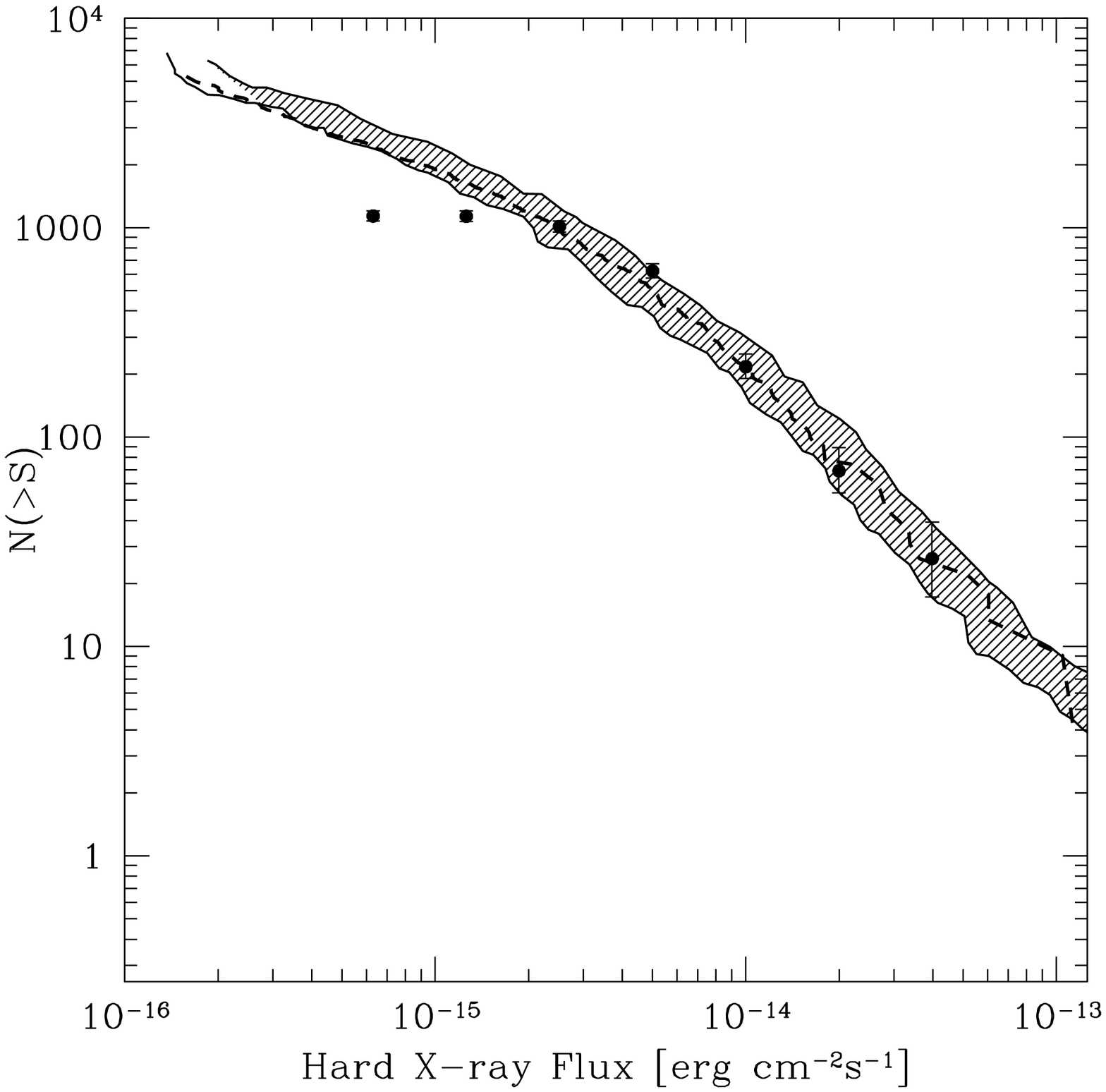}
\caption{Cumulative flux distributions for the soft (\textit{left panel}) 
and hard (\textit{right panel}) bands. (\textit{Filled circles:}) present
data for the ECDFS catalog, with error bars corresponding to the 84\% 
confidence level \citep{gehrels86}. Note that the error bars
are not independent. For comparison, we show the $\log$~N--$\log$~S compiled
by \citeauthor{moretti03} (\citeyear{moretti03}; from ROSAT, ASCA, XMM, and \chandra\ observations) with
$\pm$1$\sigma$ errors (\textit{hatched region}), and the distribution for 
sources
in the \chandra\ Deep Fields North and South (\citealp{bauer04}; 
\textit{dashed line}). 
The 2-10~keV fluxes of \citet{moretti03}
and \citet{bauer04} were converted to 2-8~keV fluxes using a
factor of 0.8, corresponding to a spectral slope of
1.4. The agreement is very good down to fluxes where incompleteness in the
ECDFS catalog becomes important (see Section 3.5), 
$\sim$2$\times 10^{-16}$~erg~cm$^{-2}$~s$^{-1}$ 
and
$\sim$2$\times 10^{-15}$~erg~cm$^{-2}$~s$^{-1}$,
in the soft and hard bands, respectively.
\label{lognlogs}}
\end{figure}

The cumulative distribution of sources for the soft and hard bands is shown in
Figure~\ref{lognlogs}.
Error bars for a given bin were calculated by adding in quadrature
the error bars from the previous
bin to the 84\% confidence error bars appropriate to the additional number
of sources in the present bin, following the procedure described in
\citet{gehrels86}. The observed distribution is compared
to the compilation of \citet{moretti03} and to the $\log$~N--$\log$~S 
for the \chandra\ deep fields reported by
\citet{bauer04}. In the soft band there is very good
agreement with the comparison sample in the flux range from $\sim
4\times10^{-14}$ to $2\times 10^{-16}$~erg~cm$^{-2}$~s$^{-1}$.
At the bright end, the discrepancy is not statistically significant, 
$\sim$1$\sigma$, because there are few bright X-ray sources in our
field. At fluxes below $\sim$2$\times 10^{-16}$~erg~cm$^{-2}$~s$^{-1}$, 
the observed $\log$~N--$\log$~S in the ECDFS flattens
relative to the comparison samples because of 
incompleteness near the flux
limit. Sources with soft fluxes of $\leq$2$\times
10^{-16}$~erg~cm$^{-2}$~s$^{-1}$ are only detected at the $\leq$
2$\sigma$ level, and thus not all sources
will be recovered. 

The $\log$~N--$\log$~S relation for the hard
band is shown in the right panel of Figure~\ref{lognlogs} and is
compared again with the distributions of \citet{moretti03} and
\citet{bauer04}. \citet{moretti03} used 2-10~keV instead
of 2-8~keV for the hard band. 
To convert 2-10~keV fluxes to the 2-8~keV band, we used a factor of 
0.8, corresponding to the flux conversion assuming a $\Gamma$=1.4
spectral slope. \citet{bauer04} quote 2-8 keV but appear
to have used 2-10 keV, so we also converted their fluxes by the same factor
(which reproduces their curve in Figure 4 of their paper).
As in the soft band, very good agreement
with previously reported $\log$~N--$\log$~S relations is seen for
the 4$\times 10^{-14}$ to 2$\times
10^{-15}$~erg~cm$^{-2}$~s$^{-1}$ range, and again,
incompleteness at the faint end explains the observed discrepancy.

\begin{figure}
\plotone{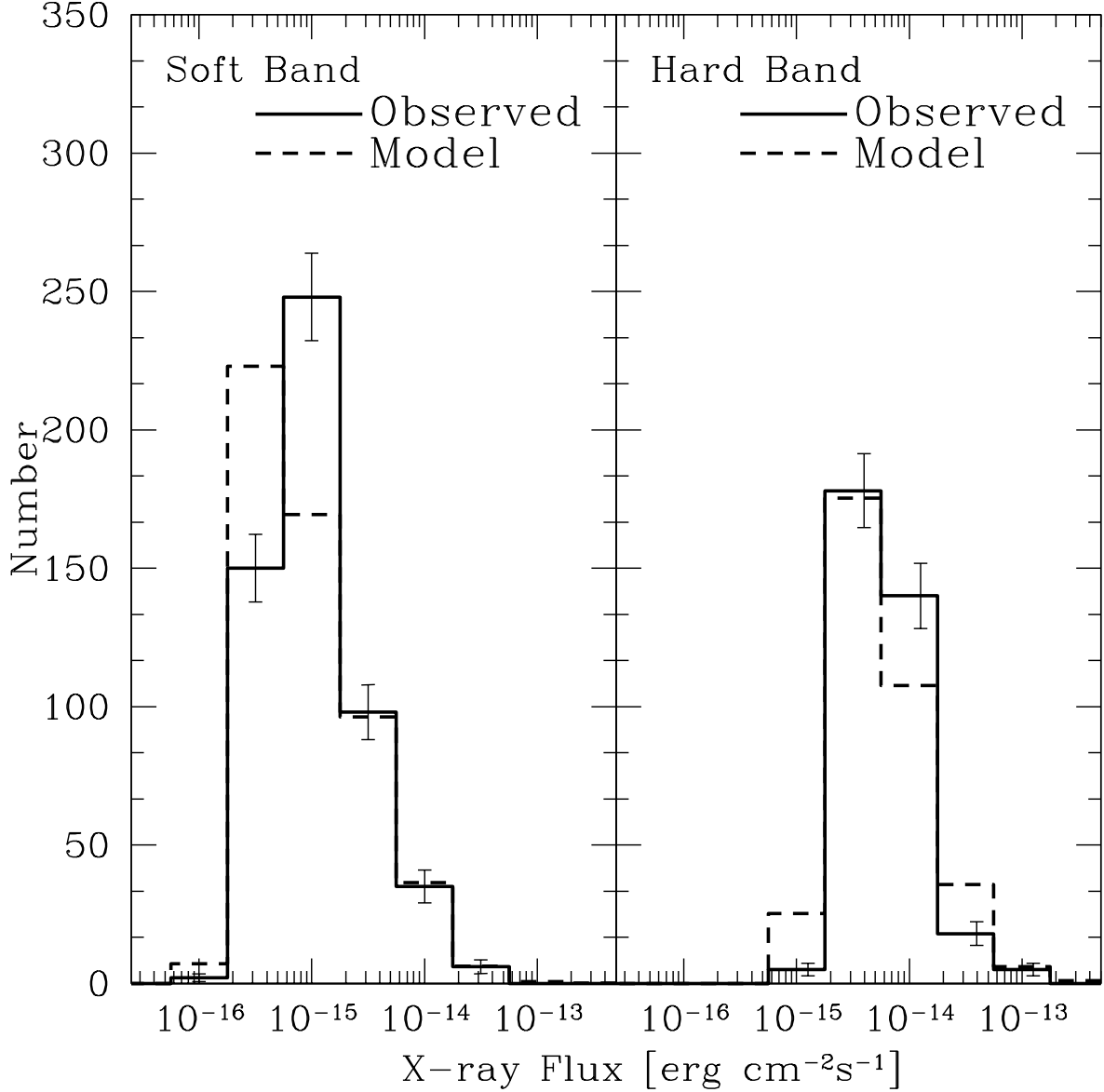}
\caption{{\it Solid lines}:  Observed differential flux distribution for 
sources in the ECDFS in the soft (left) and hard (right) 
bands, in $\Delta \log S$=0.5 bins. {\it Dashed lines}: The 
distribution predicted by an AGN unification model that also explains the X-ray
background (\citealp{treister04}, \citealp{treister05}) agrees well in the 
bright to
intermediate flux range for both bands. Below F$_X \sim$ 3 $\times$ 10$^{-16}$
~erg~cm$^{-2}$~s$^{-1}$ in the soft band, and F$_X \sim$ 1 $\times$ 10$^{-15}$
~erg~cm$^{-2}$~s$^{-1}$ in the hard band, incompleteness in our catalog 
becomes important.
\label{hist_flux}}
\end{figure}

The differential $\log$~N--$\log$~S for both the soft
and hard bands is shown in Figure~\ref{hist_flux}. These
observed distributions are compared to the predictions
of the AGN population synthesis model of \citet{treister05}
which explains the X-ray background as a superposition of mostly
obscured AGN. This model also explains the multiwavelength
number counts of AGN in the \chandra\ Deep Fields \citep{treister04}. 
Given that these models match very well to the
observed cumulative flux distributions from existing surveys, it is not
surprising that this model also successfully explains the
$\log$~N--$\log$~S distributions in the ECDFS field. 
Discrepancies can be found only at the
fainter end, where 
incompleteness causes the number of observed 
sources to fall below the model prediction.

\begin{figure}
\plotone{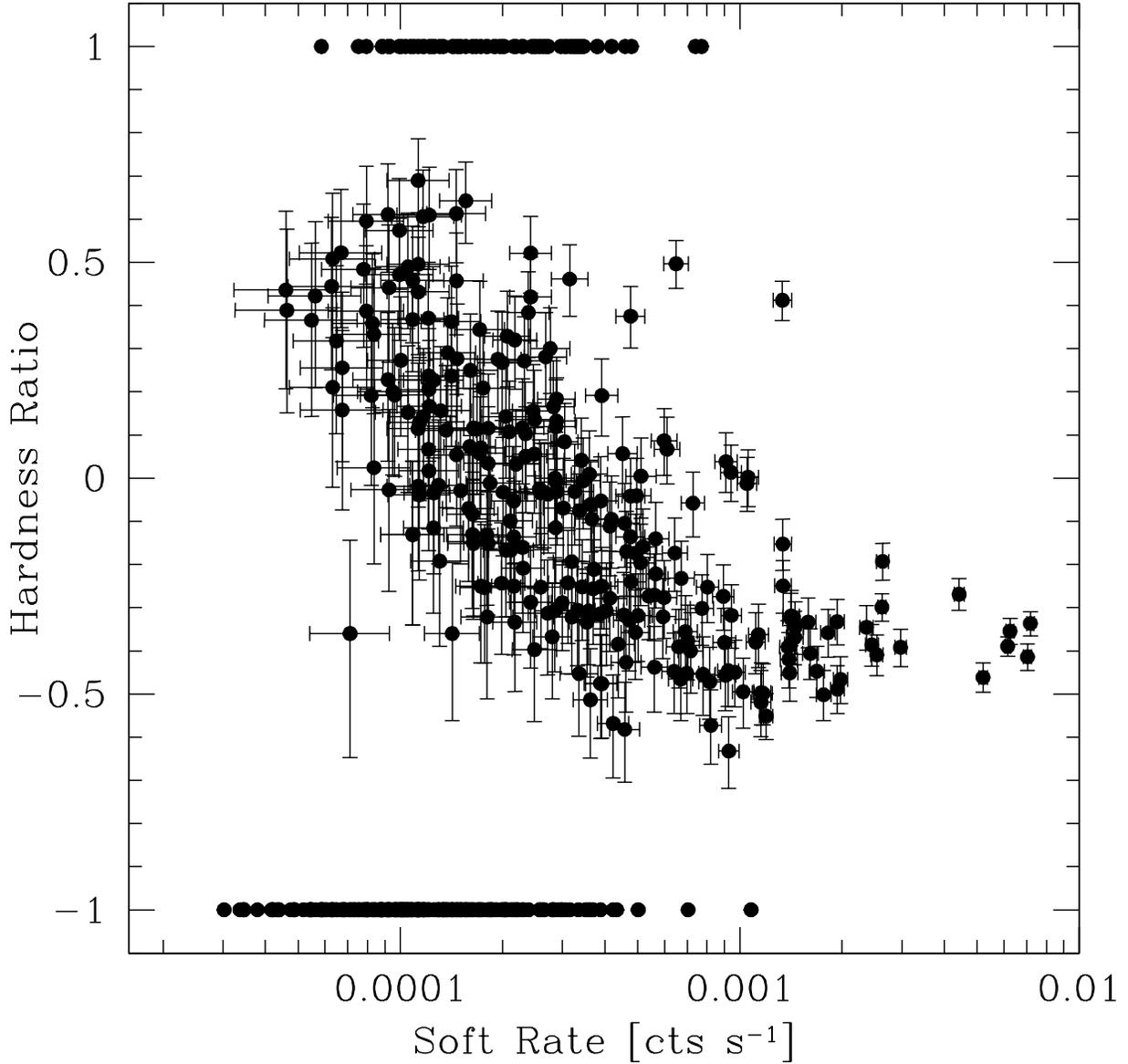}
\caption{Hardness ratio (defined as the ratio 
of hard minus soft counts to the summed counts)
versus soft X-ray count rate for sources in the
ECDFS. Error bars correspond to 84\% confidence level on
the count rates \citep{gehrels86}. 
For sources not detected in the soft band
(i.e., HR=+1), the hard count rate was used instead. Fainter
sources in the soft band have harder X-ray spectra,
supporting the hypothesis that these sources are mainly obscured
AGN, as required by population synthesis models for the X-ray
background.
\label{hr_rate}}
\end{figure}

One of the early \chandra\ results was the finding that
fainter X-ray sources have in general harder spectra
\citep{giacconi01}, represented by higher values of the
hardness ratio. Figure~\ref{hr_rate} shows that
this effect is also observed in the ECDFS field, for a much
larger number of sources. This trend is explained by obscuration since
the soft band count rate is relatively more affected than the hard band,
creating a harder observed X-ray spectrum while at the same time reducing
the observed soft flux. This is in accordance with the general picture of
AGN unification, although the precise geometry is not constrained, and it
is as expected from population synthesis models (e.g.,
\citealp{treister05} and references therein) which require a
large number of obscured AGN at moderate redshift to explain
the spectral shape of the X-ray background.

\section{Conclusions}

We present here the X-ray properties of
sources detected in deep \chandra\ observations of the
ECDFS field, the largest \chandra\ survey ever performed in terms
of both area and depth. This survey covers a total of 0.3
square degrees, roughly 3 times the area of each very deep 
\chandra\ Deep Field. A total of 651 unique sources were detected 
in the four ACIS-I pointings in this field; 81 sources were detected 
in the soft but not in the full band, while 9 were detected only in 
the hard band. Roughly 15\% of these 651 unique sources --- 118 sources
in the soft band and 73 in the hard band --- were previously detected in 
the CDF-S. The fluxes derived for these sources agree well with the fluxes
obtained from the CDF-S observations.

The X-ray $\log$~N--$\log$~S in the soft and hard bands agree
well with those derived from other X-ray surveys and with predictions
of the most recent AGN population synthesis models for the X-ray 
background.

As first discovered in early deep \chandra\ observations, we
find in this sample that faint X-ray sources have in
general harder spectra, indicating that these sources are likely
obscured AGN at moderate redshifts. This is predicted by AGN
unification models that explain the properties of the X-ray
background. A future paper will discuss the optical and near-IR properties
of these objects. This field was observed with the \textit{Spitzer} Space
Telescope by the MIPS GTO team and will also be observed by \textit{Spitzer} 
as part of an approved program related to the MUSYC survey
(PI: P. van Dokkum).

The source catalogs and images presented in this paper are
available in electronic format on the World Wide Web
(http://www.astro.yale.edu/svirani/ecdfs).  We will continue
to improve the source catalog as better calibration
information, analysis methods, and software become
available. For example, we plan to optimize the searching
for variable sources and to study the multiwavelength
properties of these X-ray sources. 

Note: After this paper was submitted, another catalog paper by \cite{lehmer05}
appeared on astro-ph. Our catalogs are similar but the analysis assumptions are
different and therefore the source catalogs differ, as do the papers. We expect
the comparison to be useful.

\acknowledgments

We thank the referee for helpful comments that improved the manuscript
and are grateful to Samantha Stevenson of the CXC
Help Desk for her help and patience in answering our many
questions regarding CIAO-related tools. We also acknowledge the help
of Jeffrey Van Duyne in cross-correlating the X-ray and optical positions.
This work was supported in part
by NASA grant HST-GO-09425.13-A. ET would like to
thank the support of Fundaci\'on Andes, Centro de
Astrof\'{\i}sica FONDAP and the Sigma-Xi foundation through
a Grant in-aid of Research. EG acknowledges support by the National Science Foundation under Grant No. AST-0201667, an NSF Astronomy and Astrophysics Postdoctoral Fellowship (AAPF).

\clearpage


\clearpage

\clearpage


\clearpage

\begin{deluxetable}{lccccccl}
\tablewidth{0pt}
\tablecaption {Journal of \chandra\ Observations of the ECDFS \label{tbl-diary}}
\scriptsize
\tablehead{
\colhead{Obs.}                                 &
\colhead{Obs.}                                 &
\multicolumn{2}{c}{Exposure Time (ks)}         &
\multicolumn{2}{c}{Aim Point}                  &
\colhead{Roll Angle}                           &
\colhead{CCDs}                         \\ 
\colhead{ID}                                 &
\colhead{Start}                              &
\colhead{Raw}                                &
\colhead{Filtered}                           &
\colhead{$\alpha_{\rm 2000}$}                &
\colhead{$\delta_{\rm 2000}$}                &
\colhead{(degrees)}                          &
\colhead{Clocked}                                   
}
\startdata
5015 & 29 Feb 2004 & 162.9 & 154.8 & 03 33 05.61 & -27 41 08.84 & 270.2 & I0--I3\\
5016 & 03 Mar 2004 &  77.2 &  76.8 & 03 33 05.61 & -27 41 08.88 & 270.2 & I0--I3\\
5017 & 14 May 2004 & 155.4 & 135.2 & 03 31 51.43 & -27 41 38.80 & 181.5 & I0--I3\\
5018 & 16 May 2004 &  72.0 &  70.2 & 03 31 51.43 & -27 41 38.79 & 181.5 & I0--I3\\
5019 & 17 Nov 2004 & 163.1 & 162.3 & 03 31 49.94 & -27 57 14.56 &   0.2 & I0--I3,S2\\
5020 & 15 Nov 2004 &  77.6 &  76.9 & 03 31 49.94 & -27 57 14.56 &   0.2 & I0--I3,S2\\
5021 & 13 Nov 2004 &  97.8 &  94.2 & 03 33 02.93 & -27 57 16.08 &   0.2 & I0--I3,S2\\
5022 & 15 Nov 2004 &  79.1 &  75.8 & 03 33 02.94 & -27 57 16.07 &   0.2 & I0--I3,S2\\
6164 & 20 Nov 2004 &  69.1 &  67.3 & 03 33 02.93 & -27 57 16.04 &   0.2 & I0--I3,S2\\

\enddata
\end{deluxetable}

\clearpage
\begin{deluxetable}{llcc}
\tabletypesize{\small}
\tablewidth{0pt}
\tablecaption {Exposure Time Per Pointing \label{table-exp}}
\tablehead{
\colhead{Pointing}                            &                                
\colhead{Obsids}                              & 
\colhead{Raw Exposure}                        &
\colhead{Net Exposure}                        \\
\colhead{}                                    &
\colhead{}                                    &
\colhead{(ks)}                                &
\colhead{(ks)}
}
\startdata
Northeast  & 5015, 5016       & 240.1 & 231.6  \\
Northwest  & 5017, 5018       & 227.4 & 205.4  \\
Southeast  & 5021, 5022, 6164 & 240.7 & 237.4  \\
Southwest  & 5019, 5020       & 246.0 & 239.2  \\
\hline
Mean       &                  & 238.6 & 228.4  \\
\enddata
\end{deluxetable}

\clearpage

\begin{deluxetable}{ll}
\tabletypesize{\small}
\tablewidth{0pt}
\tablecaption {Definition of Energy Bands and Hardness Ratio \label{table-defn}}
\tablehead{
\colhead{Name}                            &                                
\colhead{Definition}                          
}
\startdata
Full Band (F)        &  0.5--8.0~keV  \\
Soft Band (S)        &  0.5--2.0~keV  \\
Hard Band (H)        & 2.0--8.0~keV   \\
\hline
Hardness Ratio (HR)  &  $(H-S)/(H+S)$                             \\
\enddata
\end{deluxetable}

\clearpage


\begin{deluxetable}{llcccccccccccccccc}
\tablecolumns{18}
\tabletypesize{\tiny}
\rotate
\tablewidth{0pc}
\tablecaption{Primary Catalog of X-ray sources in the ECDFS field ($p_{thresh}$ = $1\times 10^{-7}$). \label{cat}}
\tablehead{
\colhead{ID} & \colhead{Name} & \colhead{RA} & \colhead{Dec} & \colhead{PSF} & \multicolumn{3}{c}{Count Rate: Full Band} &
\multicolumn{3}{c}{Count Rate: Soft Band} & \multicolumn{3}{c}{Count Rate: Hard Band} & \colhead{FB Flux$^{\rm a}$} & \colhead{SB Flux$^{\rm a}$} &
\colhead{HB Flux$^{\rm a}$} & \colhead{Notes} \\
\colhead{} & \colhead{CXOYECDF} & \multicolumn{2}{c}{J2000} & \colhead{$''$} & \colhead{value} & \colhead{upper} & 
\colhead{lower} & \colhead{value} & \colhead{upper} & \colhead{lower} & \colhead{value} & \colhead{upper} & \colhead{lower} &
\multicolumn{3}{c}{erg cm$^{-2}$s$^{-1}$} & \colhead{} 
}
\startdata
 1 & J033335.6-273935 & 03 33 35.56 & -27 39 35.3 & 1.75 & 7.64e-04 & 8.26e-04 & 7.07e-04 & 5.05e-04 & 5.56e-04 & 4.59e-04 & -------- & -------- & -------- & 9.16e-15 & 2.95e-15 & 6.44e-15 &\\
 2 & J033334.9-274209 & 03 33 34.93 & -27 42 08.5 & 1.64 & 9.00e-03 & 9.20e-03 & 8.80e-03 & -------- & -------- & --------  & 2.88e-03 & 3.00e-03 & 2.77e-03 & 1.08e-13 & 3.30e-14 & 6.38e-14 &\\
 3& J033332.9-274011& 03 33 32.92& -27 40 11.2& 1.45& 2.16e-04& 2.51e-04& 1.85e-04& 1.94e-04& 2.28e-04& 1.65e-04& --------& --------& --------& 2.59e-15& 1.14e-15& 1.82e-15&\\
 4& J033332.8-274908& 03 33 32.79& -27 49 08.0& 4.05& 1.81e-03& 1.91e-03& 1.73e-03& 1.06e-03& 1.13e-03& 9.95e-04& 4.45e-04& 4.93e-04& 4.01e-04& 2.17e-14& 6.21e-15& 9.83e-15&\\
 5& J033329.8-274009& 03 33 29.84& -27 40 09.1& 1.17& 3.45e-04& 3.89e-04& 3.07e-04& 1.81e-04& 2.14e-04& 1.54e-04& --------& --------& --------& 4.14e-15& 1.06e-15& 2.91e-15&

\enddata
\tablecomments{This table is published in its entirety in the electronic edition of the Journal. A portion is shown here for guidance regarding its form and content.}
\tablenotetext{a}{Flux: Corrected for Galactic absorption with N$_H$ = 9 
$\times$ 10$^{19}$ cm$^{-2}$ assuming $\Gamma$ = 1.4.}

\end{deluxetable}

\clearpage

\begin{deluxetable}{llcccccccccccccccc}
\tablecolumns{18}
\tabletypesize{\tiny}
\rotate
\tablewidth{0pc}
\tablecaption{Secondary Catalog of X-ray sources in the ECDFS field ($p_{thresh}$ = $1\times 10^{-6}$). \label{cat2}}
\tablehead{
\colhead{ID} & \colhead{Name} & \colhead{RA} & \colhead{Dec} & \colhead{PSF} & \multicolumn{3}{c}{Count Rate: Full Band} &
\multicolumn{3}{c}{Count Rate: Soft Band} & \multicolumn{3}{c}{Count Rate: Hard Band} & \colhead{FB Flux$^{\rm a}$} & \colhead{SB Flux$^{\rm a}$} &
\colhead{HB Flux$^{\rm a}$} & \colhead{Notes} \\
\colhead{} & \colhead{CXOYECDF} & \multicolumn{2}{c}{J2000} & \colhead{$''$} & \colhead{value} & \colhead{upper} & 
\colhead{lower} & \colhead{value} & \colhead{upper} & \colhead{lower} & \colhead{value} & \colhead{upper} & \colhead{lower} &
\multicolumn{3}{c}{erg cm$^{-2}$s$^{-1}$} & \colhead{} 
}
\startdata
  1 &  J033331.7-273850 &  03 33 31.70 &  -27 38 50.3 &  0.75 &  1.34e-04 &  1.63e-04 &  1.10e-04 &  -------- &  -------- &  -------- &  -------- &  -------- &  -------- &  1.61e-15 &  4.91e-16 &  1.13e-15 &  f\\
  2 &  J033329.0-274558 &  03 33 28.97 &  -27 45 58.1 &  0.93 &  2.38e-04 &  2.74e-04 &  2.06e-04 &  -------- &  -------- &  -------- &  -------- &  -------- &  -------- &  2.85e-15 &  8.70e-16 &  2.00e-15 &  f\\
  3 &  J033326.4-273522 &  03 33 26.36 &  -27 35 22.4 &  1.06 &  1.43e-04 &  1.72e-04 & 1.18e-04 &  -------- &  -------- &  -------- &  -------- &  -------- &  -------- &  1.71e-15 &  5.22e-16 &  1.20e-15  & \\
  4 &  J033323.9-273828 &  03 33 23.91 &  -27 38 27.8 &  0.47 &  1.30e-04 &  1.58e-04 &  1.06e-04 &  -------- &  -------- &  -------- &  -------- &  -------- &  -------- &  1.55e-15 &  4.75e-16 &  1.09e-15 &  f\\
  5 &  J033231.2-273919 &  03 32 31.18 &  -27 39 18.5 &  1.23 &  3.37e-04 &  3.79e-04 &  2.99e-04 &  6.48e-05 &  8.62e-05 &  4.83e-05 &  -------- &  -------- &  --------  & 4.04e-15 &  3.78e-16 &  2.84e-15  & \\

\enddata
\tablecomments{This table is published in its entirety in the electronic edition of the Journal. A portion is shown here for guidance regarding its form and content.}
\tablenotetext{a}{Flux: Corrected for Galactic absorption with N$_H$ = 9 
$\times$ 10$^{19}$ cm$^{-2}$ assuming $\Gamma$ = 1.4.}

\end{deluxetable}

\clearpage

\begin{deluxetable}{llccl}
\tablecolumns{18}
\tabletypesize{\tiny}
\tablewidth{0pc}
\tablecaption{Catalog of Collapsed \textit{wavdetect} X-ray sources in the ECDFS Survey \label{cat3}}
\tablehead{
\colhead{ID} & \colhead{Name} & \colhead{RA} & \colhead{Dec} & \colhead{Notes}\\
\colhead{} & \colhead{CXOYECDF} & \multicolumn{2}{c}{J2000} &  \colhead{} 
}
\startdata
1 &	J033208.9-275910 &	03 32 08.87 &	-27 59 10.10 &	full \\
2 &	J033203.3-280128 &	03 32 03.33 &	-28 01 27.90 &	full, hard \\
3 &	J033201.5-280004 &	03 32 01.50 &	-28 00 03.94 &	full \\
4 &	J033151.8-280035 &	03 31 51.79 &	-28 00 34.80 &	full \\
5 &	J033150.9-280154 &	03 31 50.87 & 	-28 01 53.85 &	full \\

\enddata
\tablecomments{This table is published in its entirety in the electronic edition of the Journal. A portion is shown here for guidance regarding its form and content.}

\end{deluxetable}

\clearpage

\begin{deluxetable}{lccccc}
\tablewidth{0pt}
\tablecaption {Summary of \chandra\ Source Detections \label{tbl-sourcesummary}}
\scriptsize
\tablehead{
\colhead{Energy}                                 &
\colhead{Number of}                              &
\multicolumn{4}{c}{Detected Counts Per Source}   \\
\colhead{Band}                                      &
\colhead{Sources$^{\rm a}$ }                        &
\colhead{Maximum}                                   &
\colhead{Minimum}                                   &
\colhead{Median}                                    &
\colhead{Mean}                                      
}
\startdata
Full Band      & 561 & 2403.0 & 2.9 & 55.4 & 127.6  \\
Soft Band      & 529 & 1643.5 & 4.4 & 32.4 &  89.9  \\
Hard Band      & 335 &  757.4 & 3.3 & 42.7 &  75.2  \\
\enddata
\tablenotetext{a}{There are 651 independent X-ray sources detected with 
either a false-positive probability threshold of $1\times 10^{-7}$ 
(Table~\ref{cat}) or $1\times 10^{-6}$ (Table~\ref{cat2}).}
\end{deluxetable}

\clearpage

\begin{deluxetable}{lcccc}
\tablewidth{0pt}
\tablecaption {Sources Detected in One Band But Not Another (Primary and Seconday Tables Combined)
\label{tbl-sourcesummary2}}
\scriptsize
\tablehead{
\colhead{Detection}                              &
\multicolumn{4}{c}{Non-Detection Energy Band}    \\
\colhead{Energy Band}                            &
\colhead{Full}                                   &
\colhead{Soft}                                   &
\colhead{Hard}                                   &
\colhead{Neither}
}
\startdata
Full          &  0 & 113 & 235 & 56     \\
Soft          & 81 &   0 & 260 & 81   \\
Hard          &  9 &  65 & 0   &  9  \\
\enddata
\tablecomments{For example, there were 113 sources out of 651 that are detected in the
full band but not in the soft band.}
%
\end{deluxetable}

\clearpage






\end{document}